\documentclass[11pt]{article}
\usepackage{graphicx}
\usepackage{amsmath}
\usepackage{epsfig}

\newcommand{\BABARPubYear}    {04}

\newcommand{\BABARConfNumber} {33}
\newcommand{\SLACPubNumber} {10603}

\input babarsym

\long\def\inst#1{\par\nobreak\kern 4pt\nobreak
    {\it #1}\par\vskip 10pt plus 3pt minus 3pt}

\def\Bflav {\ensuremath{B_{\text{flav}}}\xspace}
\def\Btag {\ensuremath{B_{\text{tag}}}\xspace}

\def\ppm{\ensuremath{\pm}}
\def\acp{\ensuremath{\mathcal{A_{CP}}}\xspace}
\def\spk{\ensuremath{S_{\phi K}}\xspace}
\def\cpk{\ensuremath{C_{\phi K}}\xspace}

\def\finalscb{\ensuremath{+0.50\pm 0.25\, (\mbox{\small stat})^{+0.07}_{-0.04} 
(\mbox{\small syst})} \xspace}
\def\finalccb{\ensuremath{0.00\pm 0.23\, (\mbox{\small stat}) \pm 0.05\, (\mbox{\small syst})} \xspace}
\def\finalacp{\ensuremath{0.054\pm 0.056\, (\mbox{\small stat}) \pm 0.012\, (\mbox{\small syst})} \xspace}

\begin{document}
{\pagestyle{empty}

\begin{flushright}
{\normalsize 
\babar-CONF-\BABARPubYear/\BABARConfNumber \\
SLAC-PUB-\SLACPubNumber \\
August 2004 }
\end{flushright}

\par\vskip 5cm

\begin{center}
\Large \bf Measurements of 
\boldmath{\CP} Asymmetries in the Decay \boldmath{$B\to \phi K$}
\end{center}
\bigskip

\begin{center}
\large The \babar\ Collaboration\\
\mbox{ }\\
\today
\end{center}
\bigskip \bigskip

\begin{center}
\large \bf Abstract
\end{center}
We present a preliminary measurement of the time-dependent \CP asymmetry 
for the neutral
$B$-meson decay $\Bz\to\phi K^0$. We use a sample of approximately 227 million 
$B$-meson pairs recorded at the $\Upsilon(4S)$ resonance with the \babar\ detector 
at the \pep2\ $B$-meson Factory at SLAC. We reconstruct the \CP eigenstates 
$\phi \KS$ and $\phi \KL$ where $\phi\to K^+K^-$, $\KS\to\pi^+\pi^-$, and \KL\ 
is observed via its hadronic interactions. The other $B$ meson in the event
is tagged as either a $\Bz$ or $\Bzb$ from its decay products.
The values of the \CP-violation parameters derived from the combined $\phi K^0$
dataset are $\spk = \finalscb$ and $\cpk = \finalccb$.
In addition, we measure the \CP -violating charge asymmetry 
$\acp (B^+\to \phi K^+) = \finalacp $.
All results are preliminary.

\vfill
\begin{center}

Submitted to the 32$^{\rm nd}$ International Conference on High-Energy Physics, ICHEP 04,\\
16 August---22 August 2004, Beijing, China

\end{center}

\vspace{1.0cm}
\begin{center}
{\em Stanford Linear Accelerator Center, Stanford University, 
Stanford, CA 94309} \\ \vspace{0.1cm}\hrule\vspace{0.1cm}
Work supported in part by Department of Energy contract DE-AC03-76SF00515.
\end{center}

\newpage
} 

\begin{center}
\small

The \babar\ Collaboration,
\bigskip

%
B.~Aubert,
R.~Barate,
D.~Boutigny,
F.~Couderc,
J.-M.~Gaillard,
A.~Hicheur,
Y.~Karyotakis,
J.~P.~Lees,
V.~Tisserand,
A.~Zghiche
\inst{Laboratoire de Physique des Particules, F-74941 Annecy-le-Vieux, France }
% ...

\end{center}\newpage

\section{INTRODUCTION}
\label{sec:Introduction}
Decays of $B$ mesons into charmless hadronic final states with a 
$\phi$ meson are dominated by $b\to s\bar{s}s$ gluonic penguin
amplitudes, possibly with smaller contributions from electroweak 
penguins, while other Standard Model (SM) amplitudes are strongly 
suppressed~\cite{one}. In the SM, \CP violation arises from a single 
complex phase in the Cabibbo--Kobayashi--Maskawa (CKM) quark-mixing 
matrix~\cite{ckm}. Neglecting CKM-suppressed contributions, the 
time-dependent \CP-violating asymmetries in the decays $\Bz\to \phi K^0$ 
and $\Bz\to \jpsi K^0$ are proportional to the same parameter 
$\sin 2\beta$~\cite{grossman}, where the latter decay is dominated by 
tree diagrams. Since many scenarios of physics beyond the SM introduce 
additional diagrams with heavy particles in the penguin loops and new 
\CP-violating phases, comparison of \CP-violating observables with SM 
expectations is a sensitive probe for new physics. 
Measurements of \stwob in $B$ decays to charmonium such as $\Bz\to J/\psi \KS$ 
have been reported by the \babar~\cite{sin2bnewbabar} and Belle~\cite{sin2bnewbelle}
collaborations, and the world average for \stwob is $0.731\pm 0.056$~\cite{pdg}.
In the decay $B^0\to\phi \KS$ the Belle collaboration measures 
$\stwob = -0.96\pm 0.50^{+0.09}_{-0.11}$~\cite{fu}, while the \babar\ collaboration 
(with a sample of approximately 114 million $B\bar{B}$ pairs) measures 
$\stwob = 0.47 \pm 0.34\mbox{(stat)} ^{+0.08}_{-0.06}\mbox{(syst)}$~\cite{prl} 
in the decays $B^0\to \phi\KS$ and $B^0\to \phi\KL$.

In the SM, neglecting CKM-suppressed contributions, 
the direct \CP violation in $B^+ \to \phi K^+$~\cite{charge}, 
detected as an asymmetry 
$\acp = (\Gamma_{\phi K^-} - \Gamma_{\phi K^+})/(\Gamma_{\phi K^-} + \Gamma_{\phi K^+})$ 
in the decay rates $\Gamma_{\phi K^\pm} = \Gamma(B^\pm \to \phi K^\pm)$, 
is expected to be zero; in the presence of large new-physics contributions to
the $b\to s\bar{s}s$ transition, it could be of order 1~\cite{Ciuchini:2002pd}.
The \babar\ collaboration measures (with a sample of approximately 89 million \BB pairs)
$\acp (B^{\pm} \to \phi K^{\pm}) = 0.04 \pm 0.09 \pm 0.01$~\cite{sasha}.

In this paper we report preliminary measurements of the time-dependent \CP\ 
asymmetry in the decay $B^0\to \phi K^0$ and the charge asymmetry in the decay
$B^+\to\phi K^+$ based on a sample of approximately 
227 million $B\bar{B}$ pairs collected at the $\Upsilon(4S)$ resonance 
with the \babar\ detector~\cite{Aubert:2001tu} at the \pep2\ asymmetric-energy 
\epem storage ring~\cite{pep} located at the Stanford Linear Accelerator Center. 

\section{THE BABAR DETECTOR}
The \babar\ detector is described elsewhere~\cite{Aubert:2001tu}.
The primary components used in the analysis are a charged-particle
tracking system consisting of a five-layer silicon vertex tracker
(SVT) and a 40-layer drift chamber (DCH) surrounded by a 1.5-T solenoidal
magnet with an instrumented flux return (IFR), an electromagnetic calorimeter 
(EMC) comprised of 6580 CsI(Tl) crystals, and a detector of internally reflected 
Cherenkov light (DIRC) providing excellent charged $K$ and $\pi$ 
identification~\cite{kpi} in the momentum range relevant for this analysis.

\section{ANALYSIS METHOD}
\label{sec:Analysis}
From a $\Bz\Bzb$ meson pair we fully reconstruct one meson, $B_{CP}$, in the final
state $\phi K^0$, and partially reconstruct the recoil $B$ meson, \Btag. 
We examine \Btag for evidence that it decayed either as \Bz or \Bzb (flavor tag). 
The asymmetric beam configuration in the laboratory frame provides a nominal boost
of $\beta\gamma = 0.56$ to the $\Upsilon(4S)$, which allows the determination
of the proper decay-time difference $\Delta t = t_{CP} - t_{\text{tag}}$
using the vertex separation of the two neutral $B$ mesons along the beam ($z$) axis.
The decay rate ${\text{f}}_+({\text{f}}_-)$ when the tagging meson is a $\Bz (\Bzb)$ 
is given by 
\begin{eqnarray}
{\text{f}}_\pm(\, \deltat)& = &{\frac{{\text{e}}^{{- \left| \deltat 
\right|}/\tau_{\Bz} }}{4\tau_{\Bz}}}  \, [
\ 1 \hbox to 0cm{} \mp \,\eta_f\, \spk \sin{( \deltamd  \deltat )} 
\mp \,\cpk \cos{( \deltamd  \deltat) }   ]\,, 
\label{eq:timedist}
\end{eqnarray}
where $\tau_{\Bz}$ is the neutral $B$ meson mean lifetime, 
\deltamd is the \Bz--\Bzb oscillation frequency,
and the \CP eigenvalue is $\eta_f=-1$ ($+1$) for $\phi\KS$ ($\phi\KL$).
The time-dependent \CP -violating asymmetry is defined as
$A_{\CP} \equiv ({\text{f}}_+  -  {\text{f}}_- )/
({\text{f}}_+ + {\text{f}}_- )$.
In the SM, decays that proceed purely via the $b\to s\bar{s}s$ penguin transitions have 
\CP parameters $\spk = \sin 2\beta$ and $\cpk = 0$, where $\beta \equiv \text{arg} 
\left [ -V_{cd}^{} V_{cb}^\ast / V_{td}^{} V_{tb}^\ast\right]$.
Here $V_{ik}$ is the CKM matrix element for quarks $i$ and $k$.
 
\section{EVENT RECONSTRUCTION}
The $B_{CP}$ candidate is reconstructed in the decay mode
$\phi K^0$ with $\phi\rightarrow K^+K^-$; the $K^0$ is either 
a \KL\ or a $\KS$ decaying into $\pi^+\pi^-$.
We combine pairs of oppositely charged tracks extrapolated to a 
common vertex to form $\phi$ and \KS candidates.
For the charged tracks from the $\phi$ decay we require at least 12 
measured drift-chamber coordinates and a minimal transverse momentum of 
0.1~\gevc. The tracks must also originate from within 1.5~cm of
the nominal beam spot in the plane transverse to the beam axis and 
$\pm 10$~cm along the $z$-axis.
Tracks with momentum less than 0.7~\gevc that are used to
reconstruct the $\phi$ meson are distinguished from pions
and protons via a requirement on the likelihood that 
combines \dedx information from the SVT and the DCH.
For tracks with higher momentum, \dedx in the DCH and the 
Cherenkov angle and the number of photons as measured by the 
DIRC are used in the likelihood. The two-kaon invariant mass must 
be within 15~\mevcc of the known $\phi$ mass~\cite{pdg}. 

For tracks corresponding to \KS and $B_{\rm tag}$ daughters
our requirements are less restrictive.
A $\KS\rightarrow\pi^+\pi^-$ candidate is accepted if its
two-pion invariant mass is within 15~\mevcc of the known $K^0$ mass~\cite{pdg}, 
its reconstructed decay vertex is separated from the $\phi$ decay vertex
by at least 3 standard deviations, and the projected angle 
between the line connecting the $\phi$ and \KS\ decay 
vertices and the \KS momentum direction, in the plane perpendicular
to the beam axis, is less than 45~mrad.

We identify a \KL candidate like in our $\Bz\to J/\psi \KL$ analysis~\cite{macfprd}
either as a cluster of energy deposited in the electromagnetic calorimeter 
or as a cluster of hits in two or more layers of the instrumented flux return 
that cannot be associated with any charged track in the event. 
The \KL energy is not well measured. Therefore, we determine the \KL
laboratory momentum from its flight direction as measured from the EMC or IFR
cluster, and the constraint that the invariant $\phi\KL$ mass agrees with the
known $B^0$ mass. In those cases where the \KL is detected in both the IFR and EMC 
we use the angular information from the EMC, because it has higher precision.
In order to reduce background from $\pi^0$ decays, we reject
an EMC \KL\ candidate cluster if it forms an invariant mass between 100
and 150~\mevcc with any other cluster in the event under the $\gamma\gamma$ 
hypothesis, or if it has energy greater than 1~GeV and contains two shower 
maxima consistent with two photons from a $\pi^0$ decay.  
The remaining background of \KL candidates due
to photons and overlapping showers is further reduced 
with the use of a neural network constructed from cluster shape variables, 
trained on Monte Carlo (MC) simulated $B^0\to\phi\KL$ and measured radiative 
Bhabha events, and tested on measured $e^+e^-\to\phi(\to\KS\KL )\gamma$ and 
$B^0\to J/\psi \KL$ events. 

\section{EVENT VARIABLES}
The results are extracted from an extended unbinned maximum likelihood 
fit for which we parameterize the distributions of several
kinematic and topological variables for signal and background events
in terms of probability density functions (PDFs)~\cite{oldpub}.
The selection keeps loose requirements in those variables to
include ranges dominated by background, too.
The background $B$ candidates come primarily from random combinations of tracks 
produced in events of the type $e^+e^-\to q\bar{q}$, where  
$q = u,d,s,c$ (continuum). 
Background from other $B$ decay final states 
with and without charm is estimated with MC simulations.
Opposite-\CP contributions from the $K^+K^-K^0$ final state 
($K^+K^-$ S-wave) are estimated with data using a 
moment analysis method~\cite{chung} 
to be less than 6.6\% at a 95\% confidence level and are treated as 
a systematic error. 
The shapes of event variable distributions are obtained from signal 
and background MC samples and high statistics data control samples.
In many cases parameters describing these distributions are varied
in the likelihood fit.

Each $B_{CP}$ candidate is characterized by the energy difference 
$\Delta E = E_B^* - \frac{1}{2}\sqrt{s}$ and, except for $\Bz\to\phi\KL$,
the beam-energy--substituted mass 
$\mes = \sqrt{(\frac{1}{2}s + \vec{p}_0\cdot\vec{p}_B)^2/E_0^2 - p_B^2}$~\cite{Aubert:2001tu}.
The subscripts 0 and $B$ refer to the initial $\Upsilon(4S)$ and the $B_{CP}$ candidate,
respectively, and the asterisk denotes the $\Upsilon(4S)$ rest frame.
For signal events, \DeltaE is expected to peak at zero and 
\mes at the known $B$ mass. 
We require $\Delta E < 0.08$~GeV for $\Bz\to\phi\KL$, and
$|\DeltaE|<0.1$~GeV and $\mes > 5.21 \gevcc$ for $\Bz\to\phi\KS$.
The $\phi$-meson signal in the $KK$ invariant mass, 
$m_{KK}$, is described with a relativistic P-wave Breit-Wigner 
function with parameters obtained from data.
In the fit we also use the helicity angle $\theta_H$, which is defined 
as the angle between the directions of the $K^+$ and the parent $B_{CP}$
in the $K^+K^-$ rest frame. The $\cos\theta_H$ distribution 
for pseudoscalar-vector $B$ decay modes is $\cos^2\theta_H$,
and for the combinatorial background it is nearly uniform. 

In continuum events, particles appear mostly in two jets. This topology 
can be characterized with several variables computed in the 
$\Upsilon(4S)$ frame.
One such quantity is the angle $\theta_T$ between the thrust axis of the 
$B_{CP}$ candidate and the thrust axis formed from the other charged and 
neutral particles in the event. 
We also use the angle $\theta_B$ between the $B_{CP}$ momentum and 
the beam axis, and the sum of the momenta $p_i$ of the other charged and neutral 
particles in the event weighted by the Legendre polynomials $L_0(\theta_i)$
and $L_2(\theta_i)$ where $\theta_i$ is the angle between the momentum of 
particle $i$ and the thrust axis of the $B_{CP}$ candidate.
For $\Bz\to\phi\KS$ candidates, we combine these variables 
into a Fisher discriminant ${\mathcal F}$~\cite{Fisher:et}.
In this mode, background from other $B$ decays is negligible, as 
demonstrated in MC simulation studies. 

More stringent criteria must be applied to suppress backgrounds 
in the case of $\Bz\to\phi\KL$ candidates, 
and we require $|\cos\theta_T| < 0.8$ and $|\cos\theta_B|<0.85$.
We define the missing momentum $\vec{p}_{miss}$,
calculated in the laboratory frame from the sum of beam momenta
and all tracks and EMC clusters, excluding the \KL candidate.  
We require the polar angle $\theta_{miss}$ of the
missing momentum with respect to the beam direction 
to be greater than 0.3~rad. The cosine of the
angle between $\vec{p}_{miss}$ and the \KL direction, $\theta_K$, must
satisfy $\cos\theta_K > 0.6$.  In the plane transverse to the beam
direction, the difference between the missing momentum projected along the
\KL direction and the calculated \KL momentum
must be greater than $-0.75$~\gevc. In the Fisher discriminant
we replace $|\cos\theta_B|$ by the cosine of the angle 
between the missing momentum and the $K^+$ from the  $\phi$ decay.
The dominant \CP contamination is the mode $B\to\phi K^{*0}$, where 
the $K^{*0}$ decays to $\KL\pi^0$. In the likelihood fit we explicitly 
parameterize backgrounds from both charm and charmless 
$B$ decays, differently for neutral and charged $B$ mesons, as derived 
from MC simulations. 
\par
All the other tracks and clusters that are not associated with the
reconstructed $B^0 \to \phi K^0$ decay are used to form the \Btag ,
and its flavor is determined with a multivariate tagging algorithm~\cite{jpsinew}. 
The tagging efficiency $\epsilon_i$ and mistag probability $w_i$ in six hierarchical 
and mutually exclusive categories are measured from fully reconstructed 
$B^0$ decays into the $D^{(*)-}X^+\,(X^+ = \pip, \rho^+, a_1^+)$  and 
$\jpsi K^{*0}\,(K^{*0}\to\Kp\pim)$ flavor eigenstates (\Bflav sample).
The analyzing power $\sum_{i=1}^{6} \epsilon_i (1-2w_i)^2$ is $(30.5\pm 0.4)$\%. 
\par
A detailed description of the $\Delta t$ reconstruction algorithm is 
given in Ref.~\cite{macfprd}.
The $B_{CP}$ vertex resolution is determined by the $\phi$ vertex. 
The average $\deltaz$ resolution is $190\mum$ 
and is dominated by the tagging vertex in the event.
Thus, we can characterize the resolution with the much larger 
$B_{\text{flav}}$ sample, which we fit simultaneously with the \CP samples.
The amplitudes for the $B_{CP}$ asymmetries and for the $B_{\text{flav}}$ 
flavor oscillations are reduced by the same factor due to 
wrong-flavor tags. Both distributions are convoluted with a common $\Delta t$ 
resolution function. Backgrounds are accounted for by 
adding terms to the likelihood, incorporated with different assumptions 
about their $\Delta t$ evolution and resolution function~\cite{macfprd}.

\section{MAXIMUM LIKELIHOOD FIT}
\label{mlf}
\begin{figure}[ht]
\begin{center}
\begin{tabular}{cc}
\epsfig{file=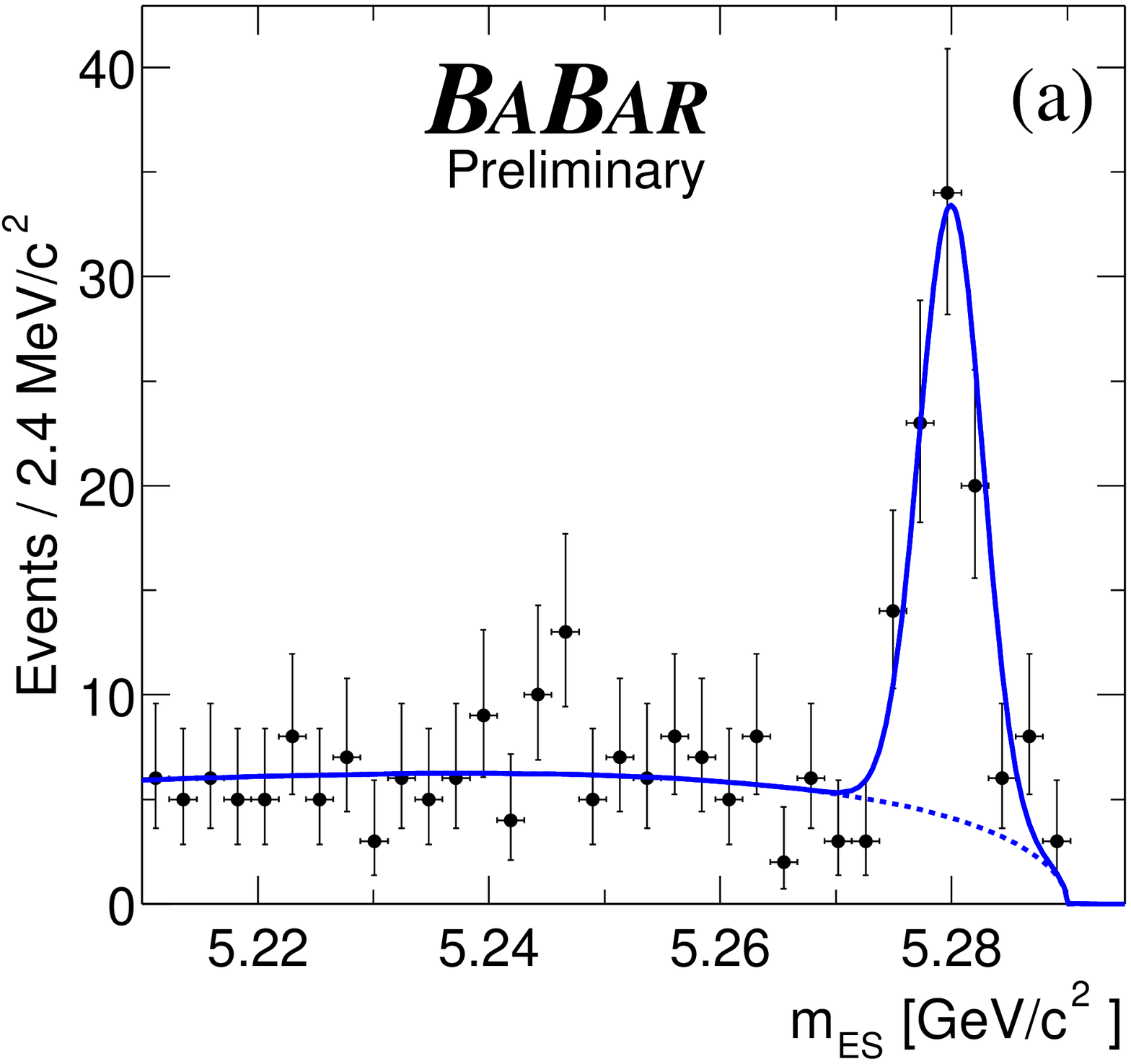,width=8.cm} &
\epsfig{file=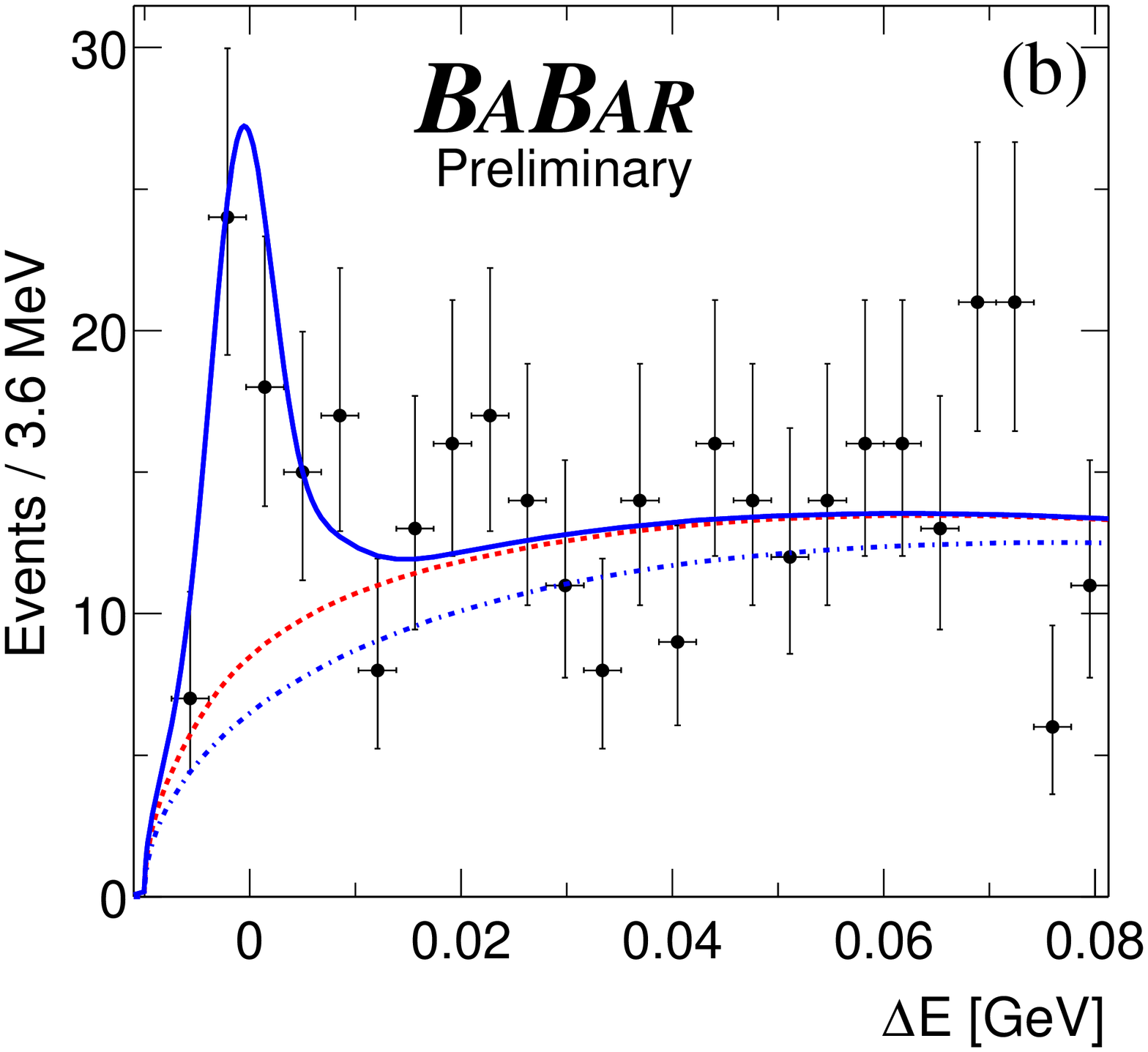,width=8.cm} \vspace{-.5cm}
\end{tabular} 
\caption{Distribution of the event variable (a) $m_{ES}$ for the 
$\phi \KS$ final state and (b) $\Delta E$ for the $\phi \KL$ final 
state after reconstruction and a requirement on the ratio of signal 
likelihood to the signal-plus-background likelihood,
calculated without the plotted variable.
The signal efficiency for the selection and likelihood requirements
is 32\% for (a) and 9\% for (b).
The solid line represents the fit result
for the total event yield and the dotted line for the total 
background. The dash-dotted (lower) line in (b) represents the 
continuum background only. \label{fig:yield}}
\end{center}
\end{figure}

Since we measure the correlations among the observables to be small in
the data samples entering the fit (the largest one is 13\% between 
$m_{ES}$ and $\Delta E$ for the signal, all others are below 7\%),
we take the probability density
function $\mathcal{P}_{i,c}^j$ for each event $j$ to be a product of the
PDFs for the separate observables. For each event hypothesis $i$
(signal, backgrounds) and tagging category $c$, for the $\phi\KS$ mode
we define
$\mathcal{P}_{i,c}^j =\mathcal{P}_i(m_{ES})\cdot\mathcal{P}_i(\Delta E)
\cdot\mathcal{P}_i(\mathcal{F})
\cdot\mathcal{P}_i(m_{KK})
\cdot\mathcal{P}_i(\cos\theta_H)
\cdot\mathcal{P}_i(\Delta t;\sigma_{\Delta t}, c)$,
for the $\phi\KL$ mode 
$\mathcal{P}_{i,c}^j =\mathcal{P}_i(\Delta E)
\cdot\mathcal{P}_i(\mathcal{F})
\cdot\mathcal{P}_i(m_{KK})
\cdot\mathcal{P}_i(\cos\theta_H)
\cdot\mathcal{P}_i(\Delta t;\sigma_{\Delta t}, c)$,
and for the flavor sample 
$\mathcal{P}_{i,c}^j =\mathcal{P}_i(m_{ES})
\cdot\mathcal{P}_i(\Delta t;\sigma_{\Delta t}, c)$.
The $\sigma_{\Delta t}$ is the error on $\Delta t$ for a given event.
The likelihood function for each decay is then
\begin{equation}
\label{likel}
{\mathcal L} = \prod_{c}\exp{\left(-\sum_{i}N_{i,c}\right)}
\prod_{j}^{N_c}\left[\sum_{i}N_{i,c}\,{\mathcal P}_{i,c}^j \right] ,
\end{equation}
where $N_{i,c}$ is the yield of events of hypothesis $i$ determined by the
fit in category $c$, and $N_c$ is the number of category $c$ events in the sample.
The total sample consists of 135,315 $B_{\text{flav}}$, 
4300 $\phi\KS$ and 8238 $\phi\KL$ candidates. 
The reconstruction efficiency for the $\phi\KS$ mode is about 40\% and 
and for the $\phi\KL$ mode about 20\%.
From the fit we find $114\pm 12$ $\phi\KS$ and $98\pm 18$ $\phi\KL$ 
signal events. The signal yields in both the $\phi K^0$ channels
agree well with our determination of the branching fraction 
for $B^0\to\phi K^0$~\cite{sasha}.
Figure~\ref{fig:yield} shows the \mes (\DeltaE ) distribution
for $\phi \KS$ ($\phi \KL$) events together with the result from 
the fit after applying a requirement on the ratio of signal likelihood 
to the signal-plus-background likelihood (computed without 
the variable plotted) to reduce the background.

We determine the $\CP$ parameters $\spk$ and $\cpk$
along with 83 other unconstrained parameters: 
event yields in signal and background (18 parameters),
distributions of kinematic and topological variables 
for signal and background (12),
the signal efficiency per tagging category (6),
the average mistag fraction and the difference  
between \Bz and \Bzb mistags for each tagging category 
in the signal (12), and the signal \deltat resolution (17). 
The \deltat parameters for the charmless $B$ background 
are the same as for the $\phi K^0$ signal.
For the $B$ decays into charm final states we parameterize the 
\deltat resolution (3) and the mistag fractions (12).
Their parameters are shared with the \Bflav sample.
The \deltat resolution for the continuum background (3)
is kept unconstrained in the $\phi K^0$ datasets.
We fix $\tau_{\Bz}$ and $\deltamd$ to the world averages~\cite{pdg}. 
The determination of the mistag fractions and \deltat-resolution
parameters is dominated by the large \Bflav sample. 
The fit was tested with both a parameterized simulation of a large
number of data-sized experiments and a full detector simulation. 
The likelihood of our data fit agrees with the likelihoods from 
fits to the simulated data. 
The fit was also verified with our $J/\psi \KS$ and  $J/\psi \KL$ 
data samples.

As a cross check the analysis was also performed using different 
selection criteria which we describe in turn.
The invariant $K^+K^-$ mass is required to be 
within 10~MeV/c$^2$ of the known mass of the $\phi$ meson 
and is not used in the likelihood fit.
The \KS flight requirements are tightened.
The same four-category multivariate tagging algorithm 
as was used for the earlier published analysis~\cite{sin2bnewbabar} is used.
Instead of the Fisher discriminant a multivariate algorithm~\cite{neuraln}
for continuum background suppression is used, which in the
$\phi\KS$ final state combines the same four variables.
In the case of $\phi\KL$ the ingredients are $L_0$, $L_2$, 
$p_{miss}$, $\cos\theta_B$, and $\cos\theta_T$.
The algorithm is trained in the same way as the Fisher discriminant
and tested on data control samples. 
The central values of \spk and \cpk for both the cross-check analysis
and the primary analysis were hidden until the analyses were complete.
We measure values for \spk and \cpk in very close agreement  
for the $\phi\KS$ and the $\phi\KL$ sample, separately,
and for the combined samples.

\begin{figure}[ht]
\begin{center}
\begin{tabular}{cc}
\epsfig{file=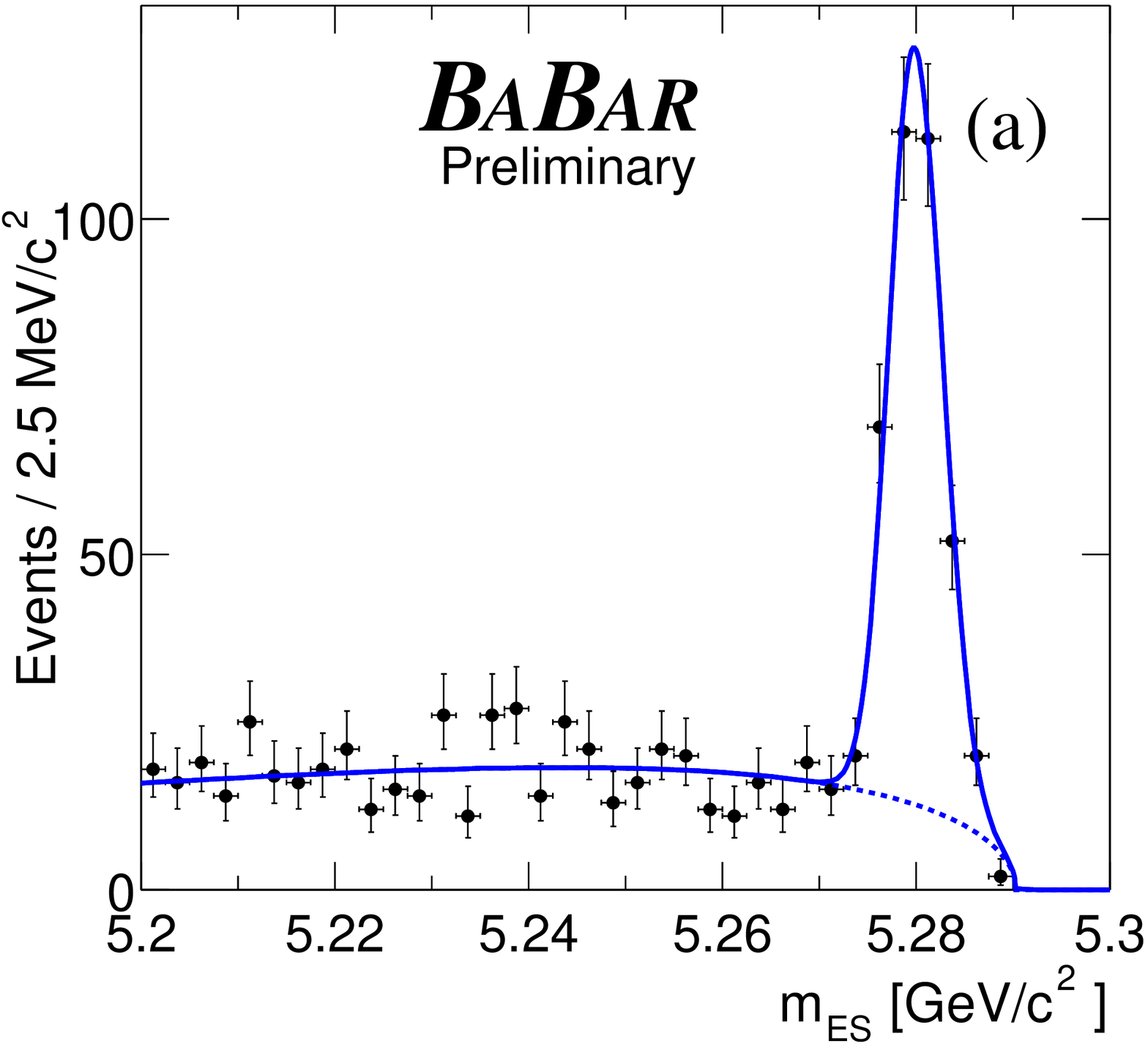,width=8.cm} &
\epsfig{file=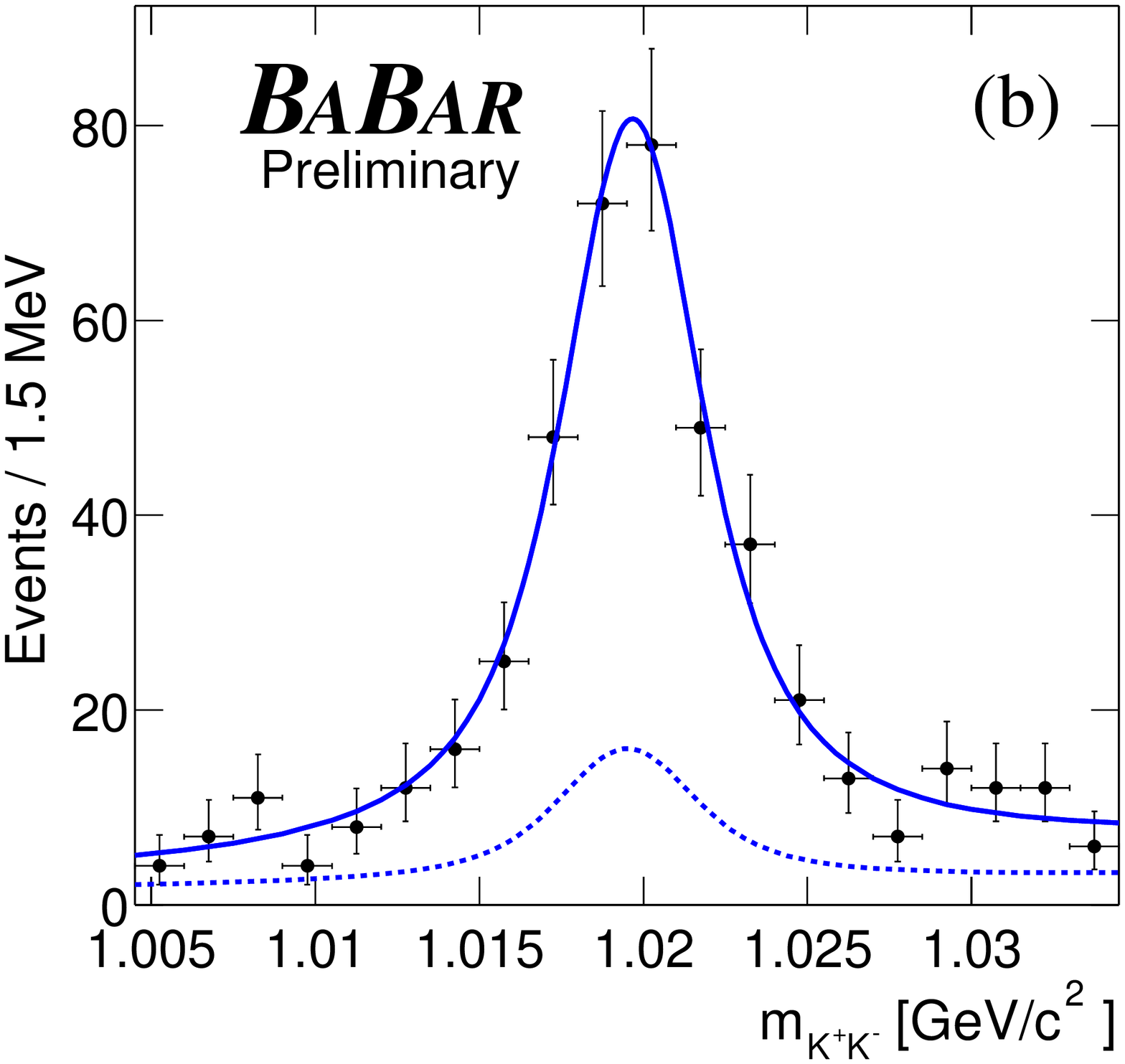,width=8.cm} \vspace{-.5cm}
\end{tabular} 
\caption{Distribution of the event variable (a) $m_{ES}$ and (b) $m_{KK}$
in the $\phi K^+$ final state after reconstruction and a requirement on
the likelihood calculated without the plotted variable. The efficiency
for the selection and likelihood requirements is 37\% for (a) and 40\%
for (b). The solid line represents the fit 
result for the total event yield and the dotted line for the background.
\label{fig:yieldkc}}
\end{center}
\end{figure}

In the measurement of the \CP -violating charge asymmetry in the decay
$B^+\to \phi K^+$ the selection of the $\phi$ meson candidate is
identical. For the $K^+$ candidate from the $B^+$ decay
the track requirements are the same as for the $\phi$ daughters 
but we apply a more restrictive
kaon identification criterion. We use the same set of event variables
as for the $\phi \KS$ channel. The likelihood is the same as in Eq.~\ref{likel}
with $c$ corresponding to the two charge categories in signal and continuum 
background. The total sample consists of 6654 $\phi K^+$ candidates and
from the fit we find $400\pm 23$ signal candidates.
Figure~\ref{fig:yieldkc} shows the distribution of
$m_{ES}$ and $m_{KK}$ with the result of the likelihood fit superimposed.
We do not observe a significant asymmetry in Monte Carlo or in the 
continuum background data.

\section{SYSTEMATIC STUDIES}
\label{sec:Systematics}
We consider systematic uncertainties in the \CP coefficients \spk and \cpk 
due to contributions from \Bz final states with opposite 
\CP (+0.06 for \spk, \ppm 0.02 for \cpk),
the parameterization of PDFs for the event yield in signal and 
background (\ppm 0.01, \ppm 0.01), 
\CP asymmetry of the background (\ppm 0.02, \ppm 0.01), 
the assumed parameterization of the $\Delta t$ resolution function (\ppm 0.02, \ppm 0.01), 
a possible difference in the efficiency for \Bz and \Bzb (\ppm 0.01, \ppm 0.02), 
the fixed values for $\Delta m_d$ and $\tau_B$ (\ppm 0.00, \ppm 0.01),
the beam-spot position (\ppm 0.01, \ppm 0.01), and
uncertainties in the SVT alignment (\ppm 0.01, \ppm 0.01).
The bias in the coefficients due to the fit procedure is included as
uncertainty (\ppm 0.01, \ppm 0.01) without making corrections to the 
final results. We estimate errors due to the effect of doubly 
CKM-suppressed decays~\cite{Long:2003wq} to be (\ppm 0.01, \ppm 0.03).
We add these contributions in quadrature to obtain the total
systematic uncertainty. 

For the measurement of the charge asymmetry \acp we estimate the uncertainty
due to charge asymmetries in tracking and particle identification to be 0.011.
We also consider the systematic error due to uncertainties in the
parameterization of the signal Fisher PDF (0.005) and $B$ background content (0.002). 
We add these contributions in quadrature to obtain the total systematic uncertainty. 

\section{RESULTS}
\label{sec:Summary}
The simultaneous fit to the $\phi K^0$ and flavor decay modes yields
the preliminary result
\begin{eqnarray}
\spk & = & \finalscb , \nonumber \\
\cpk & = & \finalccb . \nonumber
\end{eqnarray}
The preliminary results in the channel $B^0\to\phi\KS$ alone are
$\spk = 0.29\pm 0.31$ and $\cpk = -0.07\pm 0.27$, 
and in the channel $B^0\to\phi\KL$,
$\spk = 1.05\pm 0.51$ and $\cpk = 0.31\pm 0.49$,
with statistical errors only.
\begin{figure}[hpt]
\begin{center}
\vskip 1cm
\epsfig{file=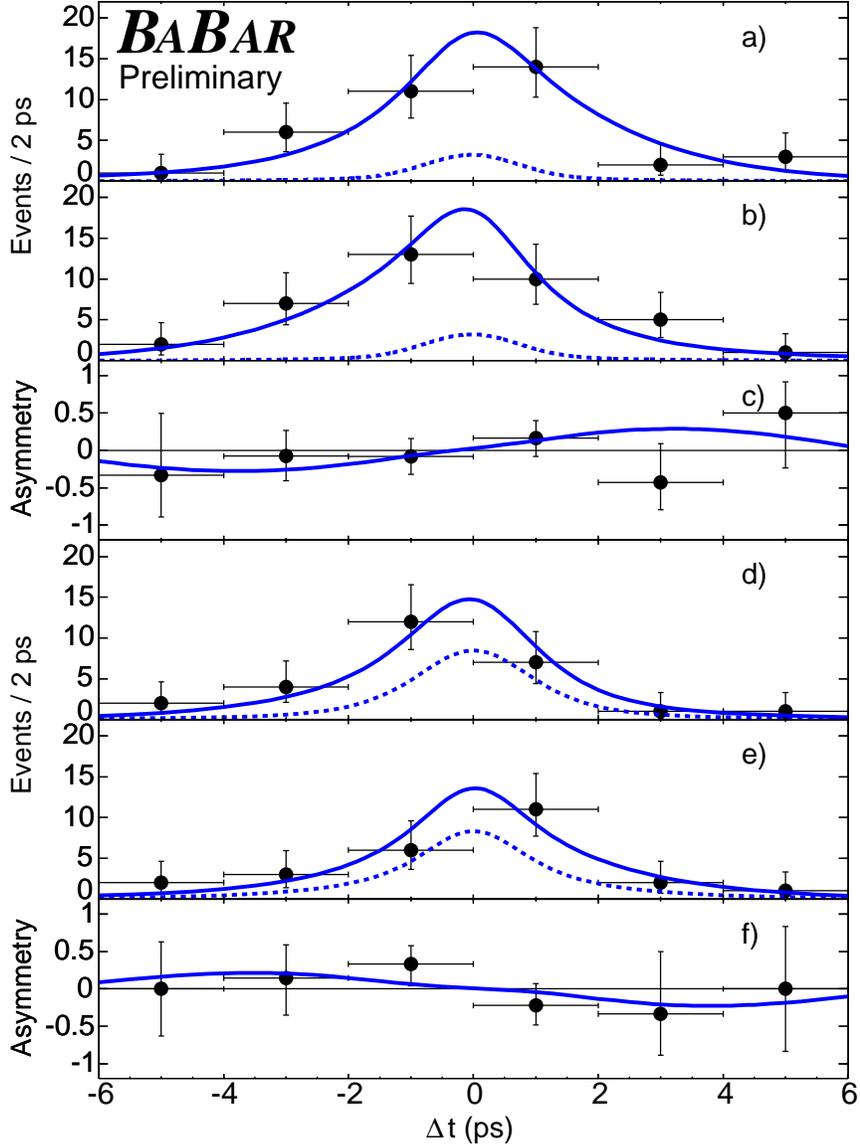, width=12cm} 
\caption{Plots a) and b) show the $\Delta t$ distributions of  
\Bz- and \Bzb-tagged $\phi\KS$ candidates. 
The solid lines refer to the fit for all events; the dashed lines 
correspond to the background. Plot c) shows the asymmetry.
Plots d), e), and f) are the corresponding plots for $\phi\KL$ 
candidates. For each final state, a requirement is applied
on the event likelihood to suppress background.
\label{fig:b0b0bar}}
\vspace*{-0.5cm}
\end{center}
\end{figure}
Figure~\ref{fig:b0b0bar} shows the $\deltat$ distributions of the \Bz- and the 
\Bzb-tagged subsets together with the raw asymmetry, for $\phi \KS$ 
and $\phi\KL$ events separately, with the result of the combined 
time-dependent \CP-asymmetry fit superimposed.

The preliminary value of the charge asymmetry in $B^+\to\phi K^+$ is
\begin{equation}
\acp = \finalacp \nonumber .
\end{equation}

\section{CONCLUSION}
\label{sec:Conclusion}
In the decay $B^0\to \phi K^0$
we measure preliminary values for \spk and \cpk
in the time-dependent \CP asymmetry that are
in close agreement with our previously published 
values~\cite{prl}. Our value of \spk agrees 
within one standard deviation with
the value of \stwob in the 
$B^0\to (\bar{c}c) K^0$ decays~\cite{jpsinew}.
We do not observe a significant
charge asymmetry in the mode $B^+\to \phi K^+$.

\section{ACKNOWLEDGMENTS}
\label{sec:Acknowledgments}
We are grateful for the 
extraordinary contributions of our \pep2\ colleagues in
achieving the excellent luminosity and machine conditions
that have made this work possible.
The success of this project also relies critically on the 
expertise and dedication of the computing organizations that 
support \babar.
The collaborating institutions wish to thank 
SLAC for its support and the kind hospitality extended to them. 
This work is supported by the
US Department of Energy
and National Science Foundation, the
Natural Sciences and Engineering Research Council (Canada),
Institute of High Energy Physics (China), the
Commissariat \`a l'Energie Atomique and
Institut National de Physique Nucl\'eaire et de Physique des Particules
(France), the
Bundesministerium f\"ur Bildung und Forschung and
Deutsche Forschungsgemeinschaft
(Germany), the
Istituto Nazionale di Fisica Nucleare (Italy),
the Foundation for Fundamental Research on Matter (The Netherlands),
the Research Council of Norway, the
Ministry of Science and Technology of the Russian Federation, and the
Particle Physics and Astronomy Research Council (United Kingdom). 
Individuals have received support from 
CONACyT (Mexico),
the A. P. Sloan Foundation, 
the Research Corporation,
and the Alexander von Humboldt Foundation.


\begin{thebibliography}{99}
\bibitem{one}
N.~G.~Deshpande and J.~Trampetic, \jprd{41}, 895 (1990);
N.~G.~Deshpande and G.~He, \plb{336}, 471 (1994);
R.~Fleischer, Z.\ Phys.\ C {\bfseries 62}, 81 (1994);
Y.~Grossman {\itshape et al.}, \jprd{68}, 015004 (2003).

\bibitem{ckm}
N.~Cabibbo, \jprl {\bfseries 10}, 531 (1963);
M.~Ko\-ba\-ya\-shi and T.~Maskawa, \progtp {\bfseries 49}, 652 (1973).

\bibitem{grossman}
A.~B.~Carter and A.~I.~Sanda, \jprd{23}, 1567 (1981);
I.~I.~Bigi and A.~I.~Sanda, \npb{193}, 85 (1981);
Y.~Grossman and M.~P.~Worah, \plb{395}, 241 (1997);
R.~Fleischer, Int.\ J.\ Mod.\ Phys.\ A {\bfseries 12}, 2459 (1997);
D.~London and A.~Soni, \plb{407}, 61 (1997).

\bibitem{sin2bnewbabar}
\babar\ Collaboration, B.~Aubert {\itshape et al.}, \jprl{89}, 201802 (2002).

\bibitem{sin2bnewbelle}
Belle Collaboration, K.~Abe {\itshape et al.}, \jprd{66}, 071102 (2002).

\bibitem{pdg}
Particle Data Group, 
S. Eidelman {\itshape et al.}, Phys. Lett. {\bf B} 592, 1 (2004).

\bibitem{fu}
Belle Collaboration, K.~Abe {\itshape et al.}, \jprl{91}, 261602 (2003).

\bibitem{prl}
\babar\ Collaboration, B.~Aubert {\itshape et al.}, \jprl{93}, 071801 (2004).

\bibitem{charge}
Charge-conjugate states are included unless explicitly stated otherwise.

\bibitem{Ciuchini:2002pd}
M.~Ciuchini and L.~Silvestrini, \jprl{89}, 231802 (2002).

\bibitem{sasha}
\babar\ Collaboration, B.~Aubert {\itshape et al.}, \jprd{69}, 011102 (2004).

\bibitem{Aubert:2001tu}
\babar\ Collaboration, B.~Aubert {\itshape et al.}, \nima{479}, 1 (2002).

\bibitem{pep}
PEP-II Conceptual Design Report, SLAC-R-418 (1993).

\bibitem{kpi}
\babar\ Collaboration, B.~Aubert {\itshape et al.}, hep-ex/0407057,
{\em submitted to} \jprl{}

\bibitem{macfprd}
\babar\ Collaboration, B.~Aubert {\itshape et al.}, \jprd{66}, 032003 (2002). 

\bibitem{oldpub}
\babar\ Collaboration, B.~Aubert {\itshape et al.}, \jprl{87}, 151801 (2001).

\bibitem{chung}
S.U.~Chung, \jprd{56}, 7299 (1997).

\bibitem{Fisher:et}
R.~A.~Fisher, Annals Eugen.\  {\bf 7}, 179 (1936).

\bibitem{jpsinew}
\babar\ Collaboration, B.~Aubert {\itshape et al.},
`Improved Measurement of Time-Dependent \CP Violation in 
$B^0\to (\bar{c}c) K^0$ Decays', BABAR-PUB-04/38,
Contribution to the 32$^{\rm nd}$ International Conference on High-Energy Physics, 
ICHEP~04.

\bibitem{neuraln}
K.S.~Cranmer, ALEPH 99-144 (1999), hep-ex/0011057.

\bibitem{Long:2003wq}
O.~Long, M.~Baak, R.~N.~Cahn and D.~Kirkby, \jprd{68}, 034010 (2003).

\end{thebibliography}
\end{document}